\documentclass[pra,twocolumn,amsmath,amssymb,groupedaddress,superscriptaddress]{revtex4-1}
\usepackage{graphics,color}
\usepackage{amsmath}
\usepackage{amssymb,mathtools}
\usepackage[dvips]{graphicx}
\usepackage{float,graphicx}
\usepackage{subfigure,hyperref}
\usepackage{epstopdf}
\usepackage{epstopdf}
\usepackage{soul}
\usepackage{epstopdf}
\newcommand{\beq}{\begin{equation}}

\newcommand{\eeq}{\end{equation}}
\newcommand{\beqa}{\begin{eqnarray}}

\newcommand{\eeqa}{\end{eqnarray}}
\newcommand{\unit}{1\!\!1}

\def\s{\sigma}

\newcommand{\bra}[1]{\langle #1 \vert} 
\newcommand{\ket}[1]{\vert #1 \rangle} 

\def\be{\begin{equation}}

\def\ee{\end{equation}}

\def\ttr{\textmd{tr}}

\def\eeq{\textmd{eq}}

\def\Rre{\textmd{Re}}

\def\Ttr{\textmd{Tr}}

\def\be{\begin{equation}}
\def\ee{\end{equation}}
\def\bea{\begin{eqnarray}}
\def\eea{\end{eqnarray}}

\usepackage{bm}
\usepackage{graphicx}

\begin{document}
\title{The quantum dynamical map of the spin boson model}
\author{In\'es de Vega}
\affiliation{Department of Physics and Arnold Sommerfeld Center for Theoretical
Physics, Ludwig-Maximilians-University Munich, Germany}

\begin{abstract}

One of the main frameworks to analyze the effects of the environment in a quantum computer is that of pure dephasing, where the dynamics of qubits can be characterised in terms of a well-known dynamical map. In this work we present a non-peturbative extension of such map beyond this simple pure-dephasing case, i.e. that is valid for a general spin coupled to a bosonic environment in a thermal state. To this aim, we use a Trotter decomposition and a Magnus expansion to simplify the unitary evolution operator in interaction picture. The proposed derivation can be extended to other finite-level open quantum systems including many body, initial system-environment correlated states, multiple-time correlation functions or quantum information protocols.  
\end{abstract}

\maketitle

\section{I. Introduction}
In the era of quantum technologies, understanding and controlling the effects of the environment in quantum mechanical systems has became a crucial task \cite{breuerbook,rivas2011a,devega2015c}. State of the art quantum computation, for instance, has been referred as \textit{noisy}-intermediate-scale quantum, to describe the fact that, currently, we can not prevent the environment from affecting the computation \cite{preskill2018}. 
This problem is described with a Hamiltonian of the form $H=H_S+H_E+H_I$, which includes the system (for instance the qubit) and environment terms, $H_S$ and $H_E$ respectively, and the interaction $H_I=\sum_\alpha S_\alpha\otimes B_\alpha$, where $S_\alpha$ and $B_\alpha$ act on the system and the environment Hilbert space, respectively. 
Despite environments can be of very different nature, Feynman and Vernon suggested in 1963 \cite{feynman1963a} that most of them can be described in terms of (real or virtual) harmonic oscillators, characterised by creation and annihilation operators $b_k^\dagger,b_k$. Hence, considering for simplicity a single component in $H_I$ we find
\bea
H_E=\sum_k \omega_k b_k^\dagger b_k;\,\,\,\, H_I=S\otimes B,
\label{HE}
\eea
with $B=\sum_k {g}_k(b_k^\dagger+b_k)$, where $g_k$ is the coupling constant of the system with each of the $k$ oscillators having a frequency $\omega_k$. With this consideration, the effects of the environment in the open system dynamics can be encoded in second order fluctuations of these operators, $C_{B}(t)=\ttr_E\{{B}(t){B}(0)\rho_E\}$, or the related spectral density, $J(\omega)$, which depend only on $g_k$ and $\omega_k$. 

Despite such a relatively simple description of the problem, obtaining the dynamics of the open quantum system (OQS) for all parameter regimes is still a challenging task. 
Simply accessing the reduced density matrix to compute its quantum mean values is in general not trivial. This quantity is obtained by tracing out the environment degrees of freedom from the total density matrix, $\rho_S(t)=\Ttr_E\{\rho(t)\}$, either numerically or analytically, and can also be written as $\rho_S(t)=\phi_t[\rho_S(0)]$ in terms of the dynamical map $\phi_t$. Interestingly, the formal structure of both the master equation, that evolves $\rho_S(t)$ \cite{dario2010,hall2014}, and the dynamical map \cite{diosi2014,vacchini2016,smirne2015} is well known. Their specific form in terms of the Hamiltonian parameters is, however, only known in certain cases: when the dynamics is reduced to the one excitation sector \cite{Vacchini2010}, in pure dephasing ($[H_S,H_I]=0$) \cite{palma1996,yu2002,cianciaruso2017,haikka2013}, or when the full Hamiltonian is quadratic \cite{hu1992,ferialdi2016}. 

Beyond these situations, the problem becomes more involved and approximations, numerical methods and special techniques are required. Among the most successful approximations is to consider a weak coupling between the system and the environment and a related perturbative expansion \cite{breuerbook,gasbarri2018}. A unitary transformation like the polaron, can allow for a similar peturbative expansion with respect to other Hamiltonian parameters \cite{mahan1981,mccutcheon2011}. Beyond such perturbative treatments, the number of approaches that have been developed during the past few years is extremely large \cite{devega2015c}. A non-comprehensive list include numerical methods like Monte-Carlo and path integrals \cite{wallsbook} as well as other stochastic methods like the stochastic Liouville-von-Neumann equation \cite{stockburger2002,stockburger2004}, that provides an accurate description of the open system  in some parameter regimes. Other numerical methods include tensor networks \cite{strathearn2018,brenes2019,somoza2019} or chain mapping techniques combined with tensor networks \cite{scholl2011,prior2010} that can be optimized to efficiently deal with finite temperature environments \cite{verstraete2004,devega2015,devega2015b,tamascelli2019}. Alternatively, one may consider hierarchy expansions that are based on expressing the correlation function $C_{B}(t)$ as a combination of exponentials \cite{tanimura2006}, or formal derivations of the OQS dynamics based on thermodynamic principles \cite{rivas2019,alipour2019}, coherent-state representations \cite{strunz2001,suess2014,devega2014}, or on re-expressing the system-environment correlations in a specific form \cite{alipour2019b}, to mention just a few.

However, obtaining a non-perturbative and computable form of the dynamical map $\phi_t$ in terms of the Hamiltonian parameters is, to the best of our knowledge, still an open problem. In this work we derive such a map for one of the most paradigmatic examples of OQS, the spin boson model, although our method can be extended to discrete OQS with more internal levels. The derived map is based on decomposing the total evolution operator in interaction picture into segments of duration $\Delta t$, and then performing a Trotter decomposition for each of them. The Trotter error is zero in the pure dephasing limit, and beyond this limit it scales with $\Delta t^2$. For this reason, the map hereby obtained can be considered as a natural extension of the pure dephasing dynamical map, which has been of extreme importance to characterize quantum information processes and dephasing in open quantum systems (see for instance \cite{palma1996,yu2002,addis2015,sung2019,bramberger2019}).


\section{Statement of the problem}
\label{QDM}

We first introduce the formal derivation of an OQS dynamical map, which is the main object we want to access with our derivation. To this aim, we consider a decorrelated initial state, $\rho(0)=\rho_S(0) \otimes \rho_E$, where $\rho_S(0)$ is the system reduced density matrix at initial time and $\rho_E$ is an environment equilibrium state, such that $[H_E,\rho_E]=0$. For simplicity, we consider a thermal state $\rho_E=\sum_m P_m|\epsilon_m\rangle\langle\epsilon_m|$, where $|\epsilon_m\rangle$ are environment eigenstates (i.e. Fock states) and $P_m=e^{-\beta\epsilon_m}/Z_E$, with $\beta=1/K_B T$ the inverse temperature ($K_B$ the Boltzmann constant), $\epsilon_m$ the environment eigenvalues, and $Z_E=\sum_m e^{-\beta\epsilon_m}$. Thus, the reduced density matrix can be written as 
\bea
\rho_S(t)&=&\Ttr_E\{U_I(t)\rho_S(0)\otimes\rho_EU_I^{\dagger}(t)\}\cr
&=&
\sum_{nm} P_m \langle \epsilon_{n}|U_I(t)|\epsilon_m\rangle \rho_S(0) \langle \epsilon_m|U_I^{\dagger}(t)|\epsilon_{n}\rangle\cr
&=&
\sum_{nm} {\mathcal E}_{nm}(t)\rho_S(0){\mathcal E}_{nm}^\dagger(t)=\phi_t[\rho_S(0)],
\label{map1}
\eea
where $U_I(t)=U_I(0,t)$ is the total evolution operator in the interaction picture with respect to the environment, which can be written as 
\bea
U_I(t)=Te^{-i\int_0^t ds H(s)},
\label{unitary}
\eea
where $T$ represents the time ordering operator, and $H(t)=H_S+H_I(t)$, with $H_I(t)=e^{iH_Et}H_Ie^{-iH_Et}$, represents the full Hamiltonian rotated with respect to $H_E$.
In addition, we have defined the Kraus operators ${\mathcal E}_{nm}=\sqrt{P_m}\langle \epsilon_{n} |U_I(t)|\epsilon_m\rangle$, which fulfil the property 
\bea
\sum_{n,m} {\mathcal E}_{nm}^\dagger {\mathcal E}_{nm}=\unit_S,
\label{property}
\eea
where $\unit_S$ is the unit operator in the system Hilbert space of dimension $d$. Very often, the Kraus decomposition (\ref{map1}) is written in terms of single label $l\equiv (n,m)$ to simplify, such that 
\bea
\rho_S(t)=\sum_{l} {\mathcal E}_{l}(t)\rho_S(0){\mathcal E}_{l}^\dagger(t)=\phi_t[\rho_S(0)].
\label{krausd}
\eea
Moreover, if the open system is a discrete-variable system (e.g. a spin), it can be described with a Gell-Mann complete basis of observables $\{G_u\,;\,u=1,\cdots,d^2-1\}$, where $d$ is the dimension of the open system (for $d=2$, they correspond to the Pauli matrices). In this case, we can write an alternative form for the map  
\bea
\rho_S(t)&=&\sum_{uv}\sum_l \Ttr_S\{{\mathcal E}_l(t)G_u\}\Ttr_S\{{\mathcal E}_{l}^\dagger(t)G_v\}G_u\rho_S(0)G_v\cr
&=&\sum_{uv}f_{uv}(t)G_u\rho_S(0)G_v,
\label{repres}
\eea
where we have defined 
\bea
f_{uv}(t)&=&\sum_l\Ttr_S\{{\mathcal E}_l(t)G_u\}\Ttr_S\{{\mathcal E}_{l}^\dagger(t)G_v\}.
\label{coeficients_2}
\eea
The map can be written in a matrix representation, where it acts over the reduced density matrix written as  vector as 
\bea
\rho_s^v(t)=\Phi_t \rho_s^v(0),
\label{mapa0}
\eea
where $\rho_s^v=(\rho_{00},\rho_{01},\cdots,\rho_{dd})$, with $\rho_{ab}=\langle a|\rho_S(t)|b\rangle$ elements in a system basis which can be for instance $(|0\rangle,\cdots,|d\rangle)$. The main problem to derive the dynamical map $\Phi_t$ is that the unitary evolution operator (\ref{unitary}) is in general quite hard to deal with due to the time ordering factor, which implies that the trace over the environment degrees of freedom in Eq. (\ref{map1}) is also quite involved. However, in this paper we manage to do both by considering a Trotter decomposition of Eq. (\ref{unitary}), followed by a Magnus expansion. This allows us to derive a compact and numerically integrable form for the dynamical map $\Phi_{t_n}=\Phi_{n}$,  with $t_n=n\Delta t$, 
\bea
\Phi_n=\sum_{uv}f_{uv}(t_n)G_u\otimes G_v.
\label{map}
\eea
In other words, our derivation allows to find a specific form for $f_{uv}$ in terms of functions that depend on the environment frequencies $\omega_k$, couplings $g_k$ and initial state coefficients $P_m$. 
\section{Trotter decomposition of the unitary evolution operator}

To access the dynamical map of the open system,  a first step is to break the total time evolution (\ref{unitary}) into short segments, 
\bea
U_I(0,t_n)&=&U_I(t_{n-1},t_n)U_I(t_{n-2},t_{n-1})\cdots \cr
&\times&U_I(t_1,t_2)U_I(0,t_1),
\label{segm}
\eea
where $t_n=n\Delta t$. The evolution operator at each segment can be written as 
\bea
U_I(t_j,t_j+\Delta t)=Te^{-i\int^{t_j+\Delta t}_{t_j}ds(H_S+H_I(s))}.
\label{unitary0}
\eea
In the following, we consider the interesting result in \cite{huy1990,poulin2011}, which approximates the former operator as $U_I(t_j,t_j+\Delta t)\approx U_I(t_j,t_j+\Delta t)^{\textmd{TS}}$ where 
\bea
U^{\textmd{TS}}_I(t_j,t_j+\Delta t)&=& Te^{-i\int^{t_j+\Delta t}_{t_j}dsH_S}Te^{-i\int^{t_j+\Delta t}_{t_j}dsH_I(s)}\cr
&=&e^{-iH_S\Delta t}Te^{-i\int^{t_j+\Delta t}_{t_j}dsH_I(s)}.
\label{TSU}
\eea
This is a generalized Trotter-Suzuki expansion, which as shown in \cite{poulin2011} gives an error in terms of the operator norm that goes like 
\bea
||U_I(t_j,t_j+\Delta t)-U^{\textmd{TS}}_I(t_j,t_j+\Delta t)||\le c_{12}(\Delta t)^2,
\eea
where 
\bea
c_{12}=\frac{1}{(\Delta t)^2}\int_{t_j}^{t_j+\Delta t}dv\int_{t_j}^v du ||[H_S,H_I(u)]||.
\label{c12}
\eea
This reduces to the usual Trotter error for the time-independent case. Thus, the Trotter approximated expansion of Eq. (\ref{segm}) can be written as 
\bea
U_I(0,t_n)&\approx&U^{\textmd{TS}}_I(t_{n-1},t_n)U^{\textmd{TS}}_I(t_{n-2},t_{n-1})\cdots \cr
&\times&U^{\textmd{TS}}_I(t_1,t_2)U^{\textmd{TS}}_I(0,t_1),
\label{segmT}
\eea
with each piece given by Eq. (\ref{TSU}). We notice that in the pure dephasing case, $c_{12}=0$ and the Trotter error vanishes. However, away from this situation, one should consider a $\Delta t$ small enough such that the error remains bounded, and the approximation (\ref{TSU}) accurate enough to describe the physics.  

Moreover, despite Eq. (\ref{TSU}) is a more simplified version of the original unitary evolution operator, it still contains a time ordered exponential, which makes very hard to numerically evaluate the trace over the environment degrees of freedom as required in Eq. (\ref{map1}). For this reason, further manipulations are needed before this trace operation can be done. However, it is important to emphasize that none of these manipulations involve further errors. 

\subsection{The system Hamiltonian}

To proceed further, we consider that our open quantum system is a spin with a general Hamiltonian $H_S=\tilde{\Delta}\sigma_x+\omega_S\sigma_z$, where $\tilde{\Delta}$ and $\omega_S$ are two system energies, $\sigma_x=|0\rangle\langle 1|+|1\rangle\langle 0|$ and $\sigma_z=|0\rangle\langle 0|-|1\rangle\langle 1|$, with $|0\rangle,||1\rangle$ the spin computational basis. This Hamiltonian can be diagonalized as 
\bea
H_S=\sum_{\alpha=+,-}E_\alpha|\alpha\rangle\langle\alpha|, 
\label{HS}
\eea
where $|\alpha\rangle$ and $E_\alpha$ are the eigenvalues and eigenstates. Notice that the eigenstates can be written in terms of the computational basis of the spin $\{|a\rangle\}=\{|0\rangle,|1\rangle\}$ as a linear combination of the form $|\alpha\rangle=\sum_{a=0,1}c_{\alpha,a}|a\rangle$, with $c_{\alpha,a}=\langle a|\alpha\rangle$. In addition, we chose we chose the coupling operator $S=\sigma_z$. As it will be shown, the obtained map has a very appealing form, and reduces to the pure dephasing one when $\tilde{\Delta}=0$, as expected. 

\subsection{Re-expressing the two terms in Eq. (\ref{TSU})}

The first term of the decomposition (\ref{TSU}) corresponds to the system evolution part with Hamiltonian (\ref{HS}) and can be written as 
\bea
M=e^{-iH_S\Delta t}=\sum_{\alpha=\pm}|\alpha\rangle\langle \alpha|e^{-iE_\alpha\Delta t}.
\label{piece1}
\eea

The second term of Eq. (\ref{TSU}) can also be simplified, reminding that time-ordered exponentials can in general be written in terms of a Magnus expansion \cite{blanes2008,blanes2010,mahan1981}, $Te^{-i\int_{t_j}^{t_j+\Delta t}dsH_I(s)}=e^{\Omega(t_j,t_j+\Delta t)}$ with $\Omega(t_j,t_j+\Delta t)=\sum_{k=1}^{\infty}\Omega_k(t_j,t_j+\Delta t)$, and 
 $\Omega(t_j,t_j)=0$. The first terms of the series have the form \cite{blanes2010}
\bea
\Omega_1(t_j,t_j+\Delta t)&=&-i\int_{t_j}^{t_j+\Delta t} dt_1H_I(t_1),\\
\Omega_2(t_j,t_j+\Delta t)&=&-\frac{1}{2}\int_{t_j}^{t_j+\Delta t} dt_1\int_{t_j}^{t_1} dt_2[H_I(t_1),H_I(t_2)],\cr
\Omega_3(t_j,t_j+\Delta t)&=&\frac{i}{6}\int_{t_j}^{t_j+\Delta t} dt_1\int_{t_j}^{t_1} dt_2\int_{t_j}^{t_2} dt_3\cr
&\times&\bigg([H_I(t_1),[H_I(t_2),H_I(t_3)]\cr
&+&[H_I(t_3),[H_I(t_2),H_I(t_1)]\bigg).
\label{magn}
\eea
In our case, 
\bea
H_{I}(t)&=&e^{iH_Et}H_Ie^{-iH_Et}\cr
&=& \s_z \sum_k g_k \left(b_k^\dagger e^{i\omega_k t} + b_k e^{-i\omega_k t}\right) \cr
[H_{I}(t_1) , H_{I}(t_2)] &=& 
2i\sum_k g_k^2 \sin(\omega_k (t_2-t_1)).
\eea
Since the commutator $[H_I(t_i),H_I(t_j)]$ is a c-number, the Magnus decomposition is truncated to the first two terms, i.e. 
\bea
Te^{-i\int^{t_j}_{t_j+\Delta t}dsH_I(s)}=e^{\Omega_1^{(j)}+\Omega_2^{(j)}},
\label{evolution}
\eea
where we have used the compact notation $\Omega_1^{(j)}=\Omega_1 (t_j,t_j+\Delta t)$ and $\Omega_2^{(j)}=\Omega_2 (t_j,t_j+\Delta t)$. Then, according to Eq. (\ref{magn}), the only non-zero terms in the Magnus expansion are 
\bea
\Omega^{(j)}_1 &=& -i \sigma_z \tilde{\Omega}^{(j)}_1\cr
\Omega^{(j)}_2 &=& - i \sum_k \frac{g_k^2}{\omega_k^2}(\sin(\omega_k \Delta t)-\omega_k \Delta t)=\Omega_2,
\eea
where $\Omega^{(j)}_2=\Omega_2$, since it is independent from $j$, and we have defined $\tilde{\Omega}^{(j)}_1=\sum_k (\alpha^*_{kj}  b_k^\dagger +\alpha_{kj} b_k )$, with 
\bea
\alpha_{kj}&=&g_k\int^{t_j+\Delta t}_{t_j}ds e^{-i\omega_k s}\cr
&=&g_ke^{-i\omega_k t_j}\frac{e^{-i\omega_k \Delta t}-1}{-i\omega_k},
\label{alpha}
\eea
The next step consists on rewriting $\Omega^{(j)}_1$ considering that $\sigma_z=|0\rangle\langle 0|-|1\rangle\langle 1|$, 
\bea
\Omega^{(j)}_1&=& -i\sigma_z\tilde{\Omega}^{(j)}_1=
\sum_{l_j=-i,i}l_j|l_j\rangle\langle l_j|   \tilde{\Omega}^{(j)}_1, 
\eea
where each $j$ has two possible values $l_j=-i,i$ and corresponding states of the computational basis, $|l_j=-i\rangle=|0\rangle$ and $|l_j=i\rangle=|1\rangle$. 
As the exponential of a diagonal operator in the system basis, it can be written as 
\bea
e^{\Omega^{(j)}_1}=\sum_{l_j}|l_j\rangle\langle l_j|e^{l_j\tilde{\Omega}^{(j)}_1}.
\label{piece2}
\eea
Considering Eqs. (\ref{piece1}) and (\ref{piece2}), the Trotter approximated evolution operator (\ref{TSU}) can be written as 
\bea
U^{\textmd{TS}}_I(t_j,t_j+\Delta t)=e^{\Omega_2}M \sum_{l_j}|l_j\rangle\langle l_j|e^{l_j \tilde{\Omega}^{(j)}_1}.
\label{unitary0}
\eea

\subsection{Reordering the string of evolution operators}
\label{reorder}
Having written each time segment in such a convenient form, we now analyze how to simplify the product of two segments, such that we place at its left an exponential over creation operators and at its right an exponential over annihilation operators. The reasons to do this will be clear later on, when making the trace over the environment degrees of freedom. Indeed, the product of two segments is simply, 
\bea
U^{\textmd{TS}}(t_1,t_2)U^{\textmd{TS}}(0,t_1)&=&e^{\Omega_2}e^{\Omega_2}\sum_{l_0,l_1}M |l_1\rangle\langle l_1|M |l_0\rangle\langle l_0|\cr
&\times&e^{l_1\tilde{\Omega}^{(1)}_1}e^{l_0\tilde{\Omega}^{(0)}_1},
\label{U12}
\eea 
Through the Baker-Campbell-Hausdorff (BCH) formula 
\bea
e^{A+B+\frac{1}{2}[A,B]}=e^Ae^B, \,\,\textmd{when} \,\,[A,B]=c,
\label{BCH}
\eea
we can consider $A=l_1\tilde{\Omega}_1^{(1)}$, $B=l_0\tilde{\Omega}_1^{(0)}$, such that $[A,B]=l_0l_1\sum_k(\alpha^*_{k0}\alpha_{k1}-\alpha^*_{k1}\alpha_{k0})$, so that Eq. (\ref{U12}) can be rewritten as 
\bea
&&U^{\textmd{TS}}(t_1,t_2)U^{\textmd{TS}}(0,t_1)=\big(e^{\Omega_2}\big)^2\sum_{l_0,l_1}M |l_1\rangle\langle l_1|M |l_0\rangle\langle l_0|\cr
&\times&e^{\sum_{j=0,1}l_j\tilde{\Omega}^{(j)}_1}e^{\frac{1}{2}l_0l_1\sum_k (\alpha^*_{k0}\alpha_{k1}-\alpha_{k0}\alpha^*_{k1})}.
\label{eq1}
\eea
We now use again the BCH formula, but now considering that $A=\sum_{jk}l_j\alpha_{kj}^*b_k^\dagger$ and $B=\sum_{jk}l_j\alpha_{kj}b_k$, such that $[A,B]=-\sum_{jp}l_jl_p\alpha^*_{kj}\alpha_{kp}$, such that 
\bea
e^{\sum_{j=0,1}l_j\tilde{\Omega}^{(j)}_1}&=&e^{\sum_{jk}l_j(\alpha^*_{kj}b_k^\dagger+\alpha_{kj}b_k)}\\
&=&e^{\sum_{jk}l_j\alpha^*_{kj}b_k^\dagger}e^{\sum_{jk}l_j\alpha_{kj}b_k}e^{\frac{1}{2}\sum_{jpk}l_jl_p\alpha^*_{kj}\alpha_{kp}},\nonumber
\label{eq30}
\eea
where $j,p=0,1$. Replacing this in Eq. (\ref{eq1}), we find 
\bea
&&U^{\textmd{TS}}(t_1,t_2)U^{\textmd{TS}}(0,t_1)=\big(e^{\Omega_2}\big)^2\sum_{l_0,l_1}M |l_1\rangle\langle l_1|M |l_0\rangle\langle l_0|\cr
&\times&e^{\sum_{kj}l_j\alpha^*_{kj}b_k^\dagger}e^{\frac{1}{2}\sum_{kj}l_j\alpha_{kj}b_k}F(l_0,l_1),
\label{eq2}
\eea
where we have defined 
\bea
F(l_0,l_1)=e^{l_0l_1\sum_k \alpha^*_{k0}\alpha_{k1}}e^{\frac{1}{2}\sum_{k,j=0,1}l^2_j|\alpha_{kj}|^2}.
\label{simpleF}
\eea
We now extend this computation to $n$ time steps. Following first a similar procedure as in Eq. (\ref{U12}), we find that the full evolution operator (\ref{segm}) with Trotter can be written as 
\bea
U_I(0,t_n)&\approx &\big(e^{\Omega_2}\big)^n\sum_{l_0,\cdots l_{n-1}}M |l_{n-1}\rangle\langle l_{n-1}|\cdots M |l_0\rangle\langle l_0|\cr
&\times&e^{l_{n-1}\tilde{\Omega}^{(n)}_1}\cdots e^{l_1\tilde{\Omega}^{(1)}_1}.
\label{segm2}
\eea
Similarly as with the simpler case (\ref{eq1}), this expression can be written as 
\bea
&&U_I(0,t_n)\approx\big(e^{\Omega_2}\big)^n\sum_{l_0,\cdots l_{n-1}}M |l_{n-1}\rangle\langle l_{n-1}|\cdots M |l_0\rangle\langle l_0|\cr
&\times&e^{\sum^{n-1}_{j=0}l_j\tilde{\Omega}^{(j)}_1}e^{\frac{1}{2}\sum_{j<p,k}l_pl_j(\alpha^*_{kj}\alpha_{kp}-\alpha_{kj}\alpha^*_{kp})}.
\label{segm3}
\eea
Through the BCH formula we decompose $e^{\sum_{j}l_j\tilde{\Omega}^{(j)}_1}$ exactly as Eq. (\ref{eq30}), but now with 
the sums $j,p=0,\cdots,n-1$. Replacing this in Eq. (\ref{segm3}) we find the following expression 
\bea
U_I(0,t_n)&\approx& \big(e^{\Omega_2}\big)^n\sum_{l_0,\cdots l_{n-1}}e^{\sum_{j,k}l_j\alpha^*_{kj}b_k^\dagger}
e^{\sum_{j,k}l_j\alpha_{kj}b_k}\cr
&\times&F({\bf l})\Pi_{\bf l},
\label{segm4}
\eea
where we have defined the generalization of the function (\ref{simpleF}) as 
\bea
F({\bf l})&=&e^{\sum_{j<p,k}l_pl_j\alpha^*_{kj}\alpha_{kp}}e^{\frac{1}{2}\sum_{kj}l^2_j|\alpha_{kj}|^2},
\label{efe}
\eea
and the operator
\bea
\Pi_{\bf l}&=&M|l_n\rangle\langle l_{n-1}|\cdots M|l_0\rangle\langle l_0|.
\label{pi}
\eea
Notice that for any $j$ we find that $l_j^2=-1$, such that $e^{\frac{1}{2}\sum_{kj}l^2_j|\alpha_{kj}|^2}=e^{-\frac{1}{2}\sum_{kj}|\alpha_{kj}|^2}$. In  addition, we note that each ${\bf l}=(l_0,\cdots,l_{n-1})$ corresponds to a particular string of values of each $l_j$, and that the multiple sums $\sum_{l_0,\cdots l_{n-1}}=\sum_{\bf l}$ correspond to summing over all possible strings of the form $(-i,i,i,-i,\cdots,i)$, each corresponding to a different combination of values $(l_0,\cdots,l_{n-1})$ and corresponding string of projections $\Pi_{\bf l}$. Since for each $l_j$ we have two different values, the number of combinations is of the order of $2^n$.

\section{The reduced density matrix}
The reduced density matrix of the open quantum system can be written as
\bea
\rho_S(t_n)=\Ttr_E\{U_I(0,t_n)\rho(0)U_I^{-1}(0,t_n)\}.
\eea
We may consider for simplicity that $\rho(0)=\rho_E\otimes\rho_S(0)$, and that $\rho_E$ is a thermal state. 
Considering this, as well as the expression (\ref{segm4}) for the unitary evolution operator, the reduced density matrix can be written as 
\bea
\rho_S(t_n)&=&\sum_{\bf l,l'}G({\bf l},{\bf l}')
F({\bf l})F^*({\bf l}') \tilde{\rho}_S({\bf l},{\bf l'}),
\label{rdm}
\eea
where we have considered Eqs. (\ref{efe}) and further defined the operator 
\bea
\tilde{\rho}_S({\bf l},{\bf l'})&=&\Pi_{\bf l} \rho_S(0) \Pi^\dagger_{\bf l'},
\label{proj}
\eea
and the function 
\bea
G({\bf l},{\bf l}')&=&\Ttr_E\{e^{\sum_{j,k}l_j\alpha^*_{kj}b_k^\dagger}
e^{\sum_{j,k}l_j\alpha_{kj}b_k}\rho_E \cr
&\times&e^{\sum_{j,k}l'^*_j\alpha^*_{kj}b_k^\dagger}
e^{\sum_{j,k}l'^*_j\alpha_{kj}b_k}\}.
\label{gs}
\eea
We note that the dependency on the Hamiltonian parameters $\omega_k$ and $g_k$ is encoded in the constants $\alpha_{kj}$ defined in Eq. (\ref{alpha}). Furthermore, thanks to the re-orderings of the exponentials performed in Sec. \ref{reorder}, we can perform quite easily the trace over the environment degrees of freedom. This will be seen in the next section. 

\subsection{Performing the environment trace}

The environment trace in Eq. (\ref{gs}) can now be performed in many different ways. One possibility is to express the environment degrees of freedom in terms of a Bargmann coherent state basis  
$|z \rangle=|z_{1}\rangle |z_{2}\rangle\cdots | z_{k}\rangle...$, which represents a tensor product of the states of all the $k$ environmental oscillators \cite{strunz2001,devega2006}. The basis states for each oscillator are $| z_{k}\rangle=\exp(z_{k}a_{k}^{\dagger})|{\textmd{vac}}\rangle$, where $|{\textmd{vac}}\rangle$ is the vacuum state for this oscillator. In this basis, the environment initial thermal state can be written as \cite{quantumoptics}
\bea
\rho_E=\int d\mu(z_0)P_T(z_0^*,z_0)|z_0\rangle\langle z_0|,
\eea
where we used the Gaussian measure, 
\begin{eqnarray}
\int d\mu(z)=\int d^2z e^{-|z|^2}=\prod_{k} \int \frac {d^2 z_{k} }{\pi}e^{-|z_{k}|^2},
\label{Gaussianmeasure}
\end{eqnarray}
with the notation $e^{|z|^2}=e^{\sum_kz^*_{k}z_{k}}$, and considering the thermal $P$-function, $P(z^*_0,z_0)=\prod_\lambda P_T(z^*_{0k},z_{0k})$,
\bea
P_T(z^*_{0k},z_{0k})=\frac{e^{-|z_{0k}|^2/N(\omega_k)}}{N(\omega_k)},
\eea
where $N(\omega_k)=\frac{1}{e^{\beta\omega_k}-1}$ is the Bose Einstein distribution for a bosonic bath with inverse temperature $\beta=1/K_BT$, with $K_B$ the Boltzmann constant.  
Considering this basis, we can write the Eq. (\ref{gs}) as 
\bea
G({\bf l},{\bf l}')&=&\int d\mu(z_1)\int d\mu(z_0)P(z^*_0,z_0)\cr
&\times&
\langle z_1|  e^{\sum_{j,k}l_j\alpha^*_{kj}b_k^\dagger}
e^{\sum_{j,k}l_j\alpha_{kj}b_k} |z_0\rangle\cr
&\times&\langle z_0|e^{\sum_{j,k}l'^*_j\alpha^*_{kj}b_k^\dagger} e^{\sum_{j,k}l'^*_j\alpha_{kj}b_k}
 |z_1\rangle.
 \label{GF}
\eea
We now consider the fact that for Bargmann states $\langle z_{1k}|z_{0k}\rangle=e^{z_{1k}^*z_{0k}}$, and $a_\lambda|z_{1k}\rangle=z_{1k}|z_{1k}\rangle$. Therefore, we can rewrite the function (\ref{GF}) as 
\bea
G({\bf l},{\bf l}')&=&\int d\mu(z_1)\int d\mu(z_0)P(z^*_0,z_0)\cr
&\times&e^{\sum_{j,k}l_j\alpha^*_{kj}z^*_{1k}}e^{\sum_{j,k}l_j\alpha_{kj}z_{0k}} \cr
&\times &e^{\sum_{j,k}l'^*_j\alpha^*_{kj}z^*_{0k}}e^{\sum_{j,k}l'^*_j\alpha_{kj}z_{1k}}
e^{z^*_1z_0}e^{z^*_0z_1},
 \label{GF2}
\eea
where again we have used the notation $e^{|z_0|^2}=e^{\sum_k |z_{0k}|^2}$. Notice that such a simple expression is doable thanks to the arrangement of exponentials performed in Sec. \ref{reorder}. 
The interesting thing about Eq. (\ref{GF2}) is that it corresponds to a product of Gaussian integrals, that are analytically solvable. Solving first the integral in $z_{1}$ we find (see App. \ref{Appendix1})
\bea
G({\bf l},{\bf l}')&=&\int d z^2_0P(z^*_0,z_0)e^{\sum_{j,k}(l_j+l'^*_j)\alpha^*_{kj}z^*_{0k}}\cr
&\times&
e^{\sum_{j,k}(l_j+l'^*_j)\alpha_{kj}z_{0k}}e^{\sum_{jpk}l_jl'^*_p\alpha^*_{kj}\alpha_{kp}}.
 \label{GF20}
\eea
Solving now the Gaussian integrals in $z_0$ we find 
\bea
G({\bf l},{\bf l}')&=&e^{\sum_{jpk}(l_j+l'^*_j)(l_p+l'^*_p)N(\omega_k)\alpha^*_{kj}\alpha_{kp}}\cr
&\times&e^{\sum_{jpk}l_jl'^*_p\alpha^*_{kj}\alpha_{kp}}.
 \label{GF3}
\eea
Thus, the functions (\ref{efe}) and (\ref{GF3}) and the operator (\ref{proj}) give rise to a closed and well-defined form for the the reduced density matrix (\ref{rdm}), in terms of the Hamiltonian parameters $\omega_k$, $g_k$ and the initial state of the environment. We also note that the dependency over time is implicit in the sums in $j$ and $p$ over the quantities $\alpha_{kj},\alpha_{kp}$. This will be dealt with in the next section.

\subsection{Further simplification of the coefficients}
In the following we further simplify the products $\sum_k \alpha^*_{kj}\alpha_{kp}$ that appear the functions (\ref{efe}) and  (\ref{GF3}) within the reduced density matrix (\ref{rdm}). These functions can now be written as 
\bea
F({\bf l})&=&e^{\sum_{j<p}l_pl_j\beta(t_p-t_j)}e^{-\frac{1}{2}\beta(0)},\cr
G({\bf l},{\bf l}')&=&e^{\sum_{jp}(l_j+l'^*_j)(l_p+l'^*_p)\beta_T(t_p-t_j)}\cr
&\times&e^{\sum_{jp}l_jl'^*_p\beta(t_p-t_j)},
 \label{GF4}
\eea
where we have defined 
\bea
\beta(t_p-t_j)&=&\sum_k \alpha^*_{kj}\alpha_{kp}=\sum_k g_k^2 e^{-i\omega_k (t_p-t_j)}\cr
&\times&\bigg(\frac{e^{i\omega_k\Delta t}-1}{i\omega_k}\bigg)
\bigg(\frac{e^{-i\omega_k\Delta t}-1}{-i\omega_k}\bigg)\cr
&=&4\sum_k \sin^2\bigg(\frac{\omega_k\Delta t}{2}\bigg)\frac{g^2_k}{\omega_k^2}e^{-i\omega_k (t_p-t_j)},
\eea
where we have considered (\ref{alpha}) and used the fact that $\sin^2(x/2)=(1-\cos(x))/2$. 
In a similar way, we find that 
\bea
\beta_T(t_p-t_j)&=&\sum_k N(\omega_k)\alpha^*_{kj}\alpha_{kp}=\sum_k g_k^2 N(\omega_k) e^{-i\omega_k (t_p-t_j)}\cr
&\times&\bigg(\frac{e^{i\omega_k\Delta t}-1}{i\omega_k}\bigg)
\bigg(\frac{e^{-i\omega_k\Delta t}-1}{-i\omega_k}\bigg)\cr
&=&4\sum_k N(\omega_k)\sin^2\bigg(\frac{\omega_k\Delta t}{2}\bigg)\frac{g^2_k}{\omega_k^2}e^{-i\omega_k (t_p-t_j)}.
\eea
We can therefore conclude that both $\beta(t)$ and $\beta_T(t)$ indeed depend only on the time difference, and their decay is related to the decay of the environment correlation function. 
\subsection{Building the map}

Once we have the formal expression of the reduced density matrix (\ref{rdm}), i.e. 
\bea
\rho_S(t_n)&=&\sum_{\bf l,l'}G({\bf l},{\bf l}')
F({\bf l})F^*({\bf l}') \tilde{\rho}_S({\bf l},{\bf l'}),
\label{rdm22}
\eea
where we now consider Eqs. (\ref{GF4}) and further defined the operator like in Eq. (\ref{proj}), $
\tilde{\rho}_S({\bf l},{\bf l'})=\Pi_{\bf l} \rho_S(0) \Pi^\dagger_{\bf l'}$, 
we can derive the corresponding dynamical map. This is made by considering that the spin can be described in terms of the compete base of observables conformed by Pauli matrices, $G_u\in\{\unit/\sqrt{2},\sigma_x/\sqrt{2},\sigma_y/\sqrt{2},\sigma_z/\sqrt{2}\}$. 
As discussed in Sec. \ref{QDM}, we can write the map as follows  
\bea
\rho_S(t_n)&=&\sum_{uv}\sum_{{\bf l},{\bf l'}}G({\bf l},{\bf l}')F({\bf l})F^*({\bf l}') 
 \Ttr_S\{\Pi_{\bf l}G_u\}
\cr
&\times& \Ttr_S\{\Pi^\dagger_{\bf l'}G_v\}G_u\rho_S(0)G_v,
\label{repres}
\eea
where the coefficients are given by Eq. (\ref{GF4}) and $\Pi_L({\bf l})$ is given by Eq. (\ref{pi}). We now write the map $\rho_S^v(t)=\Phi_n\rho_S^v(0)$ that propagates in time the vectorized form of the reduced density matrix, $\rho_S^v=(\rho_{00},\rho_{01},\rho_{10},\rho_{11})$, with $\rho_{ab}=\langle a|\rho_S|b\rangle$ and $|n\rangle\in\{ |0\rangle,|1\rangle\}$  as Eq. (\ref{mapa0}), $
\Phi_n=\sum_{uv}f_{uv}(t_n)G_u\otimes G_v$, where now we known the specific form of the coefficients 
\bea
f_{uv}(t_n)&=&\sum_{{\bf l},{\bf l'}}G({\bf l},{\bf l}')F({\bf l})F^*({\bf l}')  \Ttr_S\{\Pi_{\bf l}G_u\}\cr
&\times&\Ttr_S\{\Pi^\dagger_{\bf l'}G_v\}.
\label{coeficients_2}
\eea
This expression, with the coefficients given once again by Eq. (\ref{GF4}) and $\Pi_L({\bf l})$ is by Eq. (\ref{pi}) is the main result of the paper.

\section{Considerations about the numerical integration}
We have obtained a closed form for the dynamical map (\ref{mapa0}) in terms of $F({\bf l})$, $G({\bf l},{\bf l}')$ and the operator string $\Pi_{\bf l}$. The complexity of the problem has been reduced to the task of computing such quantities for every combination of values for ${\bf l}$ and ${\bf l'}$. Despite the number of such combinations will grow as $2^n\times 2^n$, where $n$ is the number of time steps that we need to consider, such that $t_n=n\Delta t$, the whole process is computationally efficient, particularly in the following two limits:
\begin{itemize}
\item 
The most obvious one is one one has to consider a small number of time steps $n$. This can be made when quantity $c_{12}$ in the Trotter error, given by Eq. (\ref{c12}), is very small. A good reference is to consider that this quantity is zero in the pure dephasing limit, when $[H_S,H_I(t)]=0$, which means that the decomposition (\ref{TSU}) is exact in this case, and one can simply chose $\Delta t=t$, where $t$ is the desired evolution time. For this reason, if our system Hamiltonian is composed of two terms $H_S=\omega_s\sigma_z+\tilde{\Delta} \sigma_x$, and $\tilde{\Delta}\ll 1$, the factor $c_{12}$ given by Eq. (\ref{c12}) will be proportional to $\tilde{\Delta}$, and thus the $\Delta t$ can be chosen to be very large. In a certain sense, the present derivation represents a natural extension of the pure dephasing limit. 
\item We can also consider the proposal in \cite{cerrillo2014,buser2017} in which the set of dynamical maps of the form (\ref{mapa0}) that propagates the open system is transformed into a set of transfer tensors, $T$, such that 
\bea
\rho^v_S(t_n)=\sum_{j=0}^{n-1}T_{n,j}\rho^v_S(t_j),
\label{tensor}
\eea
where we define 
\bea
T_{n,0}=\Phi_n-\sum^{n-1}_{j=1}T_{n,j}\Phi_j.
\eea
In our case, the dynamical maps only depend on the time difference, and so does the transfer tensor $T_{n,j}=T_{n-j}$. Thus, the different transfer tensors are built as $T_1=\Phi_1$, $T_2=\Phi_2-\Phi_1\Phi_1=\Phi_2-T_1\Phi_1$, and $T_3=\Phi_3-T_1\Phi_2-T_2\Phi_1$, and so on. The construction (\ref{tensor}) shows that the reduced density matrix at a time $t_n$ depends on the reduced density matrix at previous times. In most of the problems the environment has a finite correlation or relaxation time $\tau_c$, which is approximately given by the decay of $\beta(t_1-t_2)$ and $\beta_T(t_1-t_2)$. In this cases, the dependency over the past is limited by a cutoff $K$, such that $T_{n-j}=T_p\rightarrow 0$ when $p=n-j>K$. This implies that it is enough to obtain the dynamical map with coefficients (\ref{coeficients_2}) up to a time $t_n=t_K=\tau_c$, and this is quite advantageous since very often $\tau_c$ is much smaller than the time we need to propagate the open system. 
\end{itemize}
Away from these cases the problem can be computationally more intensive, and a careful optimization of the algorithm might be required, based for instance in storing the matrix arrays $\Pi_{\bf n}$ to advance further time steps.

\section{Simple limits}

\subsection{Pure dephasing limit}
The pure dephasing limit is that in which the system Hamiltonian (\ref{HS}) is diagonal in the same basis $\{|a\rangle\}=\{|0\rangle,|1\rangle\}$ in which also the interaction Hamiltonian is diagonal, i.e. $
H_S=\sum_{a=0,1}E_a|a\rangle\langle a|$. In this case, the reduced density matrix (\ref{rdm}) is simplified as 
\bea
\rho_S(t_n)&=&\sum_{l_0}\sum_{l'_0}M_D(l_0)\rho_s(0)M_D^{-1}(l'_0)\prod_{j>0}\delta_{l_j,l_0}\prod_{j>0}\delta_{l'_j,l'_0}
\cr
&\times&G({\bf l},{\bf l}')
F({\bf l})F^*({\bf l}'),
\label{rdm2}
\eea
where we have defined 
\bea
M_D(l_0)=(|0\rangle\langle 0|e^{-iE_0 t}\delta_{l_0,-i}+|1\rangle\langle 1|e^{-iE_1 t}\delta_{l_0,i}).
\eea
Eq. (\ref{rdm2}) can be further simplified as 
\bea
\rho_S(t_n)&=&\sum_{l_0}\sum_{l'_0}M_D(l_0)\rho_s(0)M_D^{-1}(l'_0)\prod_{j>0}\delta_{l_j,l_0}\prod_{j>0}\delta_{l'_j,l'_0}
\cr
&\times&e^{-\sum_{kj}|\alpha_{kj}|^2}e^{\sum_{kjp}l_jl'^*_p\alpha^*_{kj}\alpha_{kp}}\cr
&\times&e^{\sum_{kjp}(l_j+l'^*_j)(l_p+l'^*_p)\alpha^*_{kj}\alpha_{kp}N(\omega_k)}.
\label{rdm3}
\eea
Now we consider the following cases: 

\subsubsection{Case $l_0=l_j=l'_j=l'_0=i$}
In this situation we find that 
\bea
F({\bf l})&=&e^{-\sum_{j<p,k}\alpha^*_{kj}\alpha_{kp}}e^{-\frac{1}{2}\sum_{jk}|\alpha_{kj}|^2}\cr
F^*({\bf l}')&=&e^{-\sum_{j<p,k}\alpha_{kj}\alpha^*_{kp}}e^{-\frac{1}{2}\sum_{jk}|\alpha_{kj}|^2}\cr
G({\bf l},{\bf l}')&=&e^{\sum_{jpk}\alpha^*_{kj}\alpha_{kp}},
\eea
but $e^{\sum_{kjp}\alpha^*_{kj}\alpha_{kp}}=e^{\sum_{j<p,k}(\alpha^*_{kj}\alpha_{kp}+\alpha_{kj}\alpha^*_{kp})}e^{\sum_{kj}l_j^2|\alpha_{kj}|^2}$, and therefore $F({\bf l})F^*({\bf l}')G({\bf l},{\bf l}')=1$, so that all the time dependency in Eq. (\ref{rdm3}) disappears for this case. A similar situation occurs for $l_0=l_j=l'_j=l'_0=-i$.
 \subsubsection{Case $l_0=l_j=i$, and $l'_0=l'_j=-i$}
For this case, we find that 
\bea
F({\bf l})&=&e^{-\sum_{j<p,k}\alpha^*_{kj}\alpha_{kp}}e^{-\frac{1}{2}\sum_{jk}|\alpha_{kj}|^2}\cr
F^*({\bf l}')&=&e^{-\sum_{j<p,k}\alpha_{kj}\alpha^*_{kp}}e^{-\frac{1}{2}\sum_{jk}|\alpha_{kj}|^2}\cr
G({\bf l},{\bf l}')&=&e^{-\sum_{jpk}4N(\omega_k)\alpha^*_{kj}\alpha_{kp}}e^{-\sum_{jp}\alpha^*_{kj}\alpha_{kp}},
\eea
so that we find that 
\bea
F({\bf l})F^*({\bf l}')G({\bf l},{\bf l}')=e^{-2\sum_{jpk}(2N(\omega_k)+1)\alpha^*_{kj}\alpha_{kp}}.
\label{efes}
\eea
Now, in the continuum limit, we find that 
\bea
\sum^{n-1}_{j=0} \alpha_{kj}^*=g_k\int_0^{t_n} dse^{i\omega_ks}=g_k\frac{e^{i\omega_k t_n}-1}{i\omega_k}.
\eea
Considering this, we find that Eq. (\ref{efes}) can be written as 
\bea
F({\bf l})F^*({\bf l}')G({\bf l},{\bf l}')=e^{-8\sum_{k}\frac{g_k^2}{\omega_k^2}(2N(\omega_k)+1)\sin^2\big(\frac{\omega_kt_n}{2}\big)},
\eea
where we have considered the fact that $(1-\cos(\omega_kt))/2=\sin^2(\omega_kt/2)$. The same result is obtained for $l_0=l_j=-i$ and $l'_0=l'_j=i$.

 \subsubsection{Reduced density matrix}
 
Considering all these cases, we find that Eq. (\ref{rdm3}) can be written as 
\bea
\rho_S(t)&=& \ket{1}\bra{1}\rho_0\ket{1}\bra{1} \nonumber\\
&+&\ket{0}\bra{0}\rho_0\ket{0}\bra{0} \nonumber\\
&+&\ket{1}\bra{1}\rho_0\ket{0}\bra{0} e^{-i(E_1-E_0) t}e^{-\Gamma(t)} \nonumber\\
&+& \ket{0}\bra{0}\rho_0\ket{1}\bra{1}  e^{-i (E_0-E_1) t}e^{-\Gamma(t)} \label{eq:5.2.2.final}
\eea
Where we defined the decay rate $\Gamma(t)$ as follows:
\bea
\Gamma(t) &:=&8 \sum_k  \frac{g_k^2}{\omega_k^2}\sin^2\left(\frac{\omega_k t}{2}\right)\coth\left(\frac{\beta \omega_k}{2}\right)\cr
&=&4\Rre\bigg\{\int_0^t ds\int_0^s du C_B(s-u)\bigg\},
\label{gammaharm}
\eea
where $C_B(t-s)=\Ttr_E\{{B}(t){B}(s)\rho_E\}$. This corresponds to the standard reduced density matrix undergoing pure dephasing. 

\subsection{Simple Markov limit}
Notice that there is a very simple case where the required time step $\Delta t\gg \tau_c$, where $\tau_c$ is the decay time of the functions $\beta(t)$ and $\beta_T(t)$. In this very simple case, the map is memory-less, i.e. 
\bea
F({\bf l})&=&e^{-\frac{1}{2}\beta(0)},\cr
G({\bf l},{\bf l}')&=&e^{\sum_{j}(l_j+l'^*_j)(l_j+l'^*_j)\beta_T(0)}\cr
&\times&e^{\sum_{j}l_jl'^*_j\beta(0)},
 \label{GF4_M}
\eea
and each piece of the map obeys the semi-group property, i.e. $\Phi_{2}=\Phi_1\Phi_1$, where $\Phi_1=\sum_{uv}f_{uv}(t_1)G_u\otimes G_v$ with 
\bea
f_{uv}(t_1)&=&e^{-\beta(0)}\sum_{l_0,l'_0}e^{(l_0+l'^*_0)(l_0+l'^*_0)\beta_T(0)}
e^{l_0l'^*_0\beta(0)}\cr&\times&
\Ttr_S\{M|l_0\rangle\langle l_0|G_u\}\Ttr_S\{|l'_0\rangle\langle l'_0|MG_v\}, 
\eea
where, as usual, the sums run over $l_0=-i,i$ and $l'_0=-i,i$.

\section{Different initial states, multiple-time correlation functions and quantum information protocols}

Once we have been able to express the evolution operator in interaction picture as in Eq. (\ref{segm4}) it is possible to access other dynamical quantities of the open quantum system. In detail, we can obtain the evolution of its reduced density matrix when starting from an initial state that is decorrelated between the system and the environment \cite{halimeh2017}, like for instance a statistical mixture of the form 
\begin{eqnarray}
\rho(0)=\int d\mu(z_0) {\cal J}(z_0 , z^*_0) \tilde{\rho}_S (z^*_0 ,z_0)|z_0\rangle\langle z_0|,
\label{ch1stat1}
\end{eqnarray}
where ${\cal J}(z_0 , z^*_0)$ is the statistical probability for the member $\tilde{\rho}_S (z^*_0 ,z_0)$ of the statistical ensemble. Indeed, the reduced density matrix can be written as 
\bea
\rho_S(t_n)&=&\sum_{\bf l,l'}\tilde{\rho}^G_S({\bf l},{\bf l'})
F({\bf l})F^*({\bf l}') ,
\label{rdm2}
\eea
but now with $\tilde{\rho}^G_S({\bf l},{\bf l'})$ the following operator 
\bea
\tilde{\rho}^G_S({\bf l},{\bf l'})&=&\int d\mu(z_1)\int d\mu(z_0){\cal J}(z_0 , z^*_0)\Pi_{\bf l}\tilde{\rho}_S (z^*_0 ,z_0) \Pi^\dagger_{\bf l'}\cr
&\times&
\langle z_1|  e^{\sum_{j,k}l_j\alpha^*_{kj}b_k^\dagger}
e^{\sum_{j,k}l_j\alpha_{kj}b_k} |z_0\rangle \cr
&\times&\langle z_0|e^{\sum_{j,k}l'^*_j\alpha^*_{kj}b_k^\dagger} e^{\sum_{j,k}l'^*_j\alpha_{kj}b_k}
 |z_1\rangle .
 \label{GF2_init_corr}
\eea
In a similar way, it is possible to access multiple time correlation functions, like for instance \cite{devega2006}, 
\begin{eqnarray}
&&C_{{\bf A}}({\bf t})=\Ttr\bigg\{\prod_{i=1}^N {{ U}}^{-1}_I(t_i,0)A_i{{ U}}_I(t_i,0)\rho(0)\bigg\}
\label{ch2gen2}
\end{eqnarray}
where $A_1,\cdots,A_N$ is any array of system operators, by directly replacing the expression (\ref{segm4}) for the unitary evolution operators. Measurement-like correlations like the ones considered in \cite{milz2019} can also be obtained. In the end, independently from the particular construction considered, we will have to compute the environment trace of exponentials of linear combinations of creation and annihilation operators. These can be arranged as before, i.e. using the BCH formula (\ref{BCH}), such that in the end one one is left with terms of the form $e^{\sum_k A_k b_k^\dagger}e^{\sum_k B_k b_k}$. 

Finally, we notice that in between the unitary evolution operators that characterize the interaction with the environment, one may consider unitary operations $V$ acting over the open system only, as it happens during a quantum computation protocol \cite{milz2017,milz2019}. In this case, we would still be able to determine with our procedure the resulting evolution, conditioned to such operations, 
\bea
&&\rho^{\textmd{cond}}_S(t)=\Ttr_E\{\cdots V_2 U_I(t_1,t_2)V_1U_I(0,t_1)\cr
&\times&\rho_S(0)\otimes \rho_EU^{-1}_I(0,t_1)V_1U^{-1}_I({t}_1,{t}_2)V_2\cdots\}.
\label{eqprot}
\eea
We note that for quantum information protocols the most desirable property is the divisibility property, i.e. the property such that in the decomposition $\phi(0,t_2)=\phi(t_2,t_1)\phi(t_1,0)$, the intermediate piece $\phi(t_2,t_1)$ is also a well-defined (i.e. completely positive) dynamical map. We note that here we have changed the notation of the map to specify the time interval, such that $\phi_t\equiv \phi(t,0)$. Thanks to this property, we can always rewrite the above protocol (\ref{eqprot}) as 
\bea
&&\rho^{\textmd{cond}}_S(t)=\cdots V_2\phi(t_1,t_2)V_1\phi(0,t_1)\rho_S(0),
\eea
i.e. in terms of the subsequent application of different dynamical maps $\phi(t_{n},t_{n+1})$ in between qubit operations $V_n$. An even more desirable case is that in which besides being divisible, the dynamical map is a semigroup, such that all intermediate pieces are built in the same way, since $\phi(t_{n},t_{n+1})=\phi(t_{n+1}-t_n,0)$.
\section{Conclusions}
We have derived the dynamical map corresponding to the spin-boson model by considering, in a similar way as in a tensor network time-evolution, a Trotter decomposition of the unitary evolution operator with a time step $\Delta t$. The only particularity is that we perform such decomposition for the evolution operator in the interaction picture with respect to the environment. Further, we consider a Magnus expansion for the resulting terms and analytically solve the trace over the environment degrees of freedom. The resulting map depends on the Hamiltonian parameters $\omega_k$ and $g_k$, as well as on the environment initial state, and can be numerically computed in an efficient way. Moreover, the most convenient limits are the limit close to pure dephasing (so that the trotter error is small and $\Delta t$ can be chosen to be large) and when the environment correlations are relatively short (such that the transfer tensor method can be used efficiently). 

Moreover, since the method is based on having re-expressed the unitary evolution operator, which is a basic building block of the evolution, it can be used to describe not only quantum mean values but also any other dynamical quantity, like multiple-time correlation functions \cite{devega2006,milz2019}. In addition, it can be easily extended to deal with arbitrary system-environment initial states, and situations where the environment evolution is interrupted by local operations over the open system, as it occurs during a quantum computation protocol \cite{milz2017,milz2019}.  

We note that the resulting expression for the reduced density matrix is very similar to the one obtained in \cite{strathearn2017} based on a Trotter discretization of the Feynman path integral. The advantage of the method here proposed is that while the Feynman influence functional required in path integrals is best known for Gaussian, i.e. harmonic, environments, the present method can in principle be extended beyond this case, provided that the pure dephasing pieces of the evolution, $Te^{-i\int^{t_{j+1}}_{t_j}}ds H_I(s)$, are can be properly arranged to perform the trace \cite{bramberger2019}. Moreover, as it is shown, our method allows for a straightforward access not only to the reduced density matrix but to other system dynamical quantities. 
 
\begin{acknowledgements}
The author acknowledge M.C. Ba\~nuls, M. Bramberger, K. Modi, C. Parra, and A. Smirne for interesting discussions. This research was financially supported by DFG-grant GZ: VE 993/1-1.
\end{acknowledgements}

\appendix
\section{Gaussian integrals}
\label{Appendix1}
Each Gaussian integral in $z_1$ appearing in Eq. (\ref{GF2}), i.e.
\bea
G({\bf l},{\bf l}')&=&\int d\mu(z_1)\int d\mu(z_0)P(z^*_0,z_0)\cr
&\times&e^{\sum_{j,k}l_j\alpha^*_{kj}z^*_{1k}}e^{\sum_{j,k}l_j\alpha_{kj}z_{0k}} \cr
&\times &e^{\sum_{j,k}l'^*_j\alpha^*_{kj}z^*_{0k}}e^{\sum_{j,k}l'^*_j\alpha_{kj}z_{1k}}
e^{z^*_1z_0}e^{z^*_0z_1},
 \label{GF2app}
\eea
can be decomposed as the product of Gaussian integrals for each component $k$, with the form 
\bea
\int dz_{1k}\frac{e^{-|z_{1k}|^2}}{\pi}e^{A_kz^*_{1k}}e^{B_kz_{1k}}e^{z_{0k}z^*_{1k}}e^{z^*_{0k}z_{1k}}\cr
=e^{(A_k+z_{0k})(B_k+z^*_{0k})},
\eea
where $A_k=\sum_j l_j\alpha^*_{kj}$ and $B_k=\sum_j l'^*_j\alpha_{kj}$. The integral in $z_0$ can be solved analogously, i.e. by considering it as a product of integrals for each $z_{0k}$. Indeed, we then find that 
\bea
\int dz_{0k}\frac{e^{-|z_{0k}|^2/N(\omega_k)}}{\pi N(\omega_k)}e^{A_kz^*_{0k}}e^{B_kz_{0k}}\cr
=e^{A_k B_k N(\omega_k)},
\eea
where this time $A_k=\sum_{j}(l_j+l'^*_j)\alpha^*_{kj}$ and $B_k=\sum_{j}(l_j+l'^*_j)\alpha_{kj}$
 \bibliography{/Users/ines.vega/Dropbox/Bibtexelesdrop2}

\end{document}